\documentclass[artice,aps,pra,showpacs,twocolumn,superscriptaddress]{revtex4}
\usepackage{graphicx,color}
\usepackage{amsmath,amsthm,amsfonts,amssymb,bm}
\usepackage{times}
\usepackage{epsf}
\usepackage[colorlinks={true}]{hyperref}
\usepackage[T1]{fontenc}
\usepackage[utf8]{inputenc}
\hypersetup{citecolor={blue}, filecolor={blue}, linkcolor={blue}, urlcolor={blue}}

\newcommand{\commentold}[1]{}
\DeclareMathSymbol{:}{\mathpunct}{operators}{"3A}
\usepackage[utf8]{inputenc}
\usepackage[english]{babel}
\usepackage{amsthm}
\theoremstyle{definition}

\begin{document}

\title{The effect of Homodyne-based feedback control on quantum speed limit time}
\author{S. Haseli}
\email{soroush.haseli@uut.ac.ir}
\affiliation{Faculty of Physics, Urmia University of Technology, Urmia, Iran
}


\date{\today}

\begin{abstract}
The minimal time a system requires to transform from an initial state to target state  is defined as the
quantum speed limit time. quantum speed limit time can be applied to quantify the maximum speed of the evolution of a quantum system. Quantum speed limit time is inversely related to the speed of evolution of a quantum system. That is, shorter quantum speed limit time means higher speed of quantum evolution. In this work, we study the quantum speed limit time of a two-level atom under Homodyne-based feedback control. The results show that the quantum speed limit time is decreased by increasing feedback coefficient. So, Homodyne-based feedback control can induce speedup the evolution of quantum system.
\end{abstract}

\maketitle
\section{Introduction}	
Quantum mechanics sets a bound on speed limit of the evolution of quantum systems. The minimum evolution time between two distinguishable states of a quantum system is called quantum speed limit time (QSLT). In Ref.\cite{Mandelstam}, Mandelstam and Tamm have provided a QSLT bound for closed quantum system as 
\begin{equation}\label{a1}
\tau \geq \tau_{Q S L}=\frac{\pi \hbar}{2 \Delta E},
\end{equation}  
where $\Delta E=\sqrt{\left\langle\hat{H}^{2}\right\rangle-\langle\hat{H}\rangle^{2}}$  is the variance  of energy of the initial state and $H$ is the time-independent Hamiltonian describing the dynamics of closed quantum system. The bound in Eq.\ref{a1} is called (MT) bound. In Ref. \cite{Margolus}, Margolus and Levitin have introduced another bound with following form
\begin{equation}\label{a2}
\tau \geq \tau_{Q S L}=\frac{\pi \hbar}{2 E},
\end{equation} 
where $E=\langle\hat{H}\rangle$ is the mean energy. The bound in Eq.\ref{a2} is called (ML) bound. It has been shown that the comprehensive Bound of QSLT for closed quantum system can be obtain by combining the results of MT bound and ML bound as \cite{Giovannetti}
\begin{equation}\label{a3}
\tau \geq \tau_{Q S L}=\max \left\{\frac{\pi \hbar}{2 \Delta E}, \frac{\pi \hbar}{2 E}\right\},
\end{equation}
In practice, it is almost impossible to isolate a system from its surroundings. It is therefore of particular importance to find the pervasive bound for QSLT of open quantum systems. To this end, much work has been done to define the proper QSLT for open quantum systems \cite{Taddei,Escher,Deffner,del Campo,Zhang,Xu,Mondal,Levitin,Luo,Meng,Mirkin,Campaioli,Uzdin}. Here we use the geometric approaches
based on relative purity for driving the QSLT for open quantum systems\cite{del Campo,Zhang}. There have also been many efforts for obtaining a short QSLT \cite{Luo,Min,Zhang1,Song,Wu}. It has been shown
that memory effects in non-Markovian evolution can decrease QSLT and speed up quantum evolution \cite{Luo}. In Ref. \cite{Zhang1}, the authors have used  the external classical driving to accelerate the speed of
evolution for an open system. The quantum speedup has also been investigated under dynamical decoupling
pulses \cite{Song} and the photonic-bandgap reservoir \cite{Wu}. 

In Refs. \cite{Wiseman,Wiseman1}, Wiseman and Milburn have provided an approach to control the
dynamics of quantum systems by feeding back the measurement results to the systems. Based on the different measurement schemes that exist, the quantum feedback control is divided into two categories as quantum-jump-based and Homodyne-based feedback control. In Refs.\cite{Mirrahimi,Carvalho} , these various types of quantum feedback control is used to protect entangled states against decoherence.

In this work, we will show how Homodyne-based feedback control can affect the QSLT of open quantum systems. The work is organized as follow. In Sec.\ref{sec1} give a brief introduction about the quantum speed limit for open quantum systems. In Sec.\ref{sec2} the model for Homodyne-based feedback control is introduced. The results and discussion is provided in Sec.\ref{sec3}. Finally, we provide a conclusion in Sec.\ref{sec4}
\begin{figure}[!] 
\centerline{\includegraphics[scale=0.5]{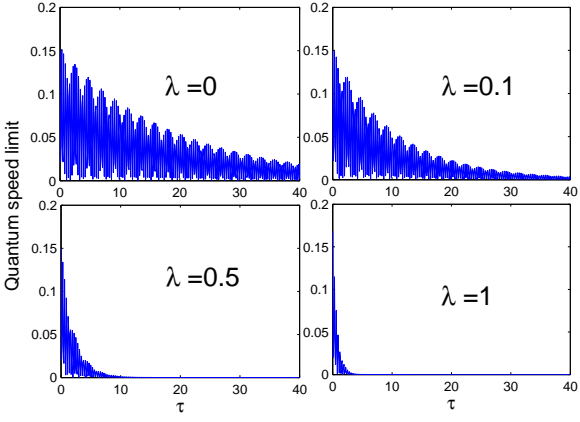}}
\vspace*{8pt}
\caption{QSLT as a function of initial time $\tau$ for different value of  feedback coefficient with $\alpha=0$, $\omega=10$, $\gamma=0.1$, $\chi=0$ and $\theta=\pi/4$.   }\label{Fig1}
\end{figure}
\begin{figure}[!] 
\centerline{\includegraphics[scale=0.5]{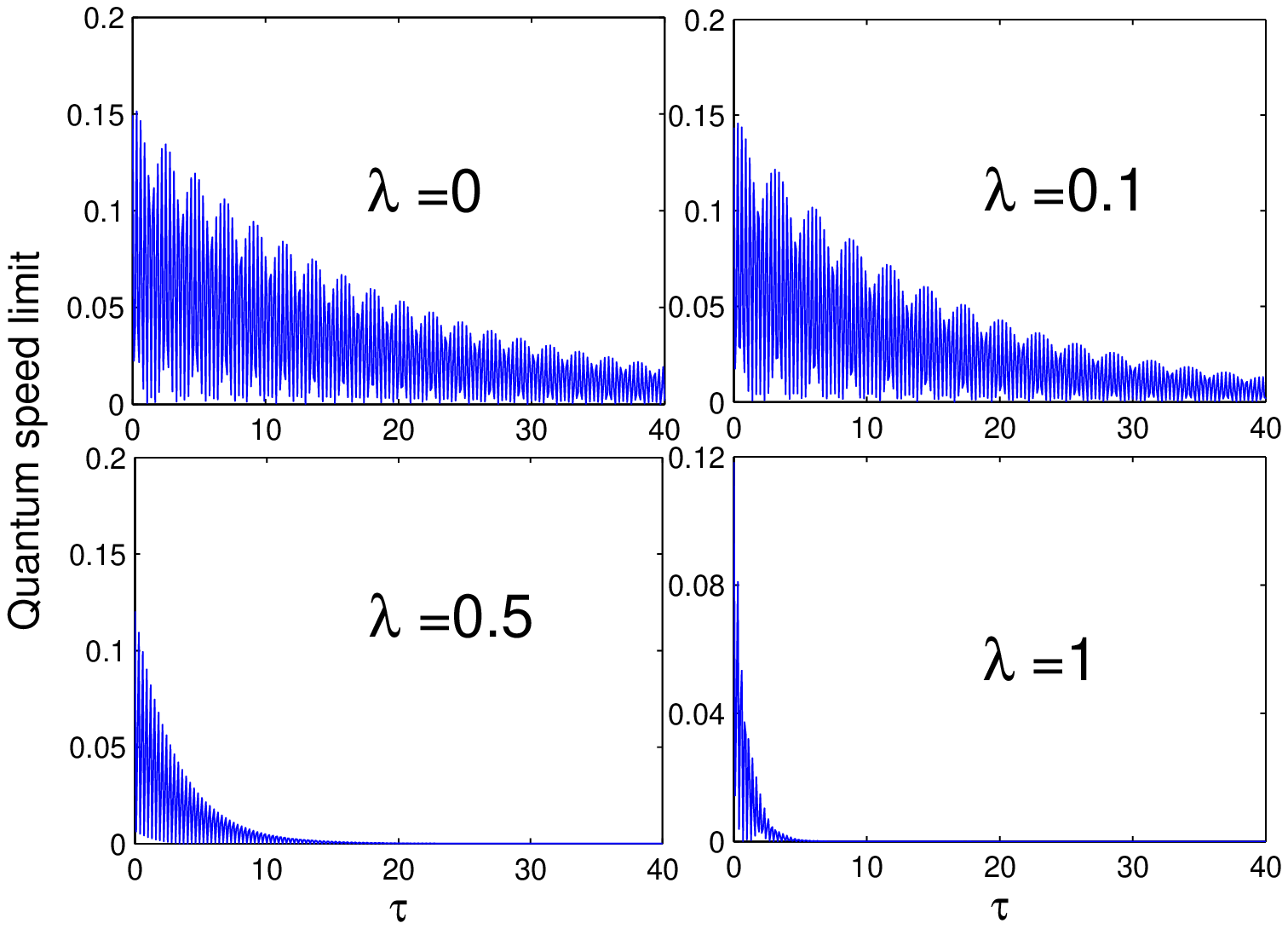}}
\vspace*{8pt}
\caption{QSLT as a function of initial time $\tau$ for different value of  feedback coefficient with $\alpha=\pi/4$, $\omega=10$, $\gamma=0.1$, $\chi=0$ and $\theta=\pi/4$. }\label{Fig2}
\end{figure}
\begin{figure}[!] 
\centerline{\includegraphics[scale=0.5]{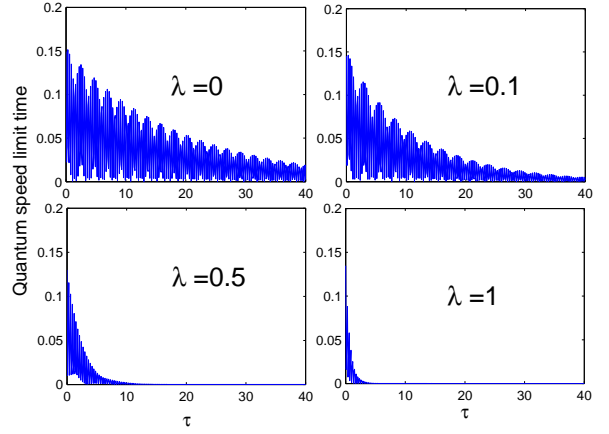}}
\vspace*{8pt}
\caption{QSLT as a function of initial time $\tau$ for different value of  feedback coefficient with $\alpha=\pi/2$, $\omega=10$, $\gamma=0.1$, $\chi=0$ and $\theta=\pi/4$.  }\label{Fig3}
\end{figure}
\section{Quantum speed limit for open quantum systems}\label{sec1}
The time evolution of an open quantum system is defined via the following time-dependent
master equation as
\begin{equation}\label{a4}
\dot{\rho}_{t}=L_{t}\left(\rho_{t}\right),
\end{equation}
where $\rho_t$ is the state of the open quantum system at time $t$ and $L_t$ is the positive generator. Quantum speed limit time is the minimum evolution time between two distinguishable states of a quantum system. Actually, QSLT is the minimum time for evolving from the state $\rho_\tau$ at time $\tau$ to target state $\rho_{\tau+\tau_D}$ at time $\tau + \tau_{D}$. Here $\tau$ is the initial time and $\tau_D$ is driving time. In Refs.\cite{del Campo,Zhang}, the authors have used relative purity to introduce QSLT for open quantum systems. They have shown that the  relative purity based QSLT can be used for arbitrary initial state. The relative purity between initial state $\rho_\tau$ and target state $\rho_{\tau+\tau_D}$ can be written as 
\begin{equation}\label{a5}
f(\tau+\tau_D)=\frac{tr(\rho_\tau\rho_{\tau+\tau_D})}{tr(\rho_\tau^{2})}.
\end{equation}
By following the calculations presented in Ref.\cite{Zhang}, one can obtain the ML bound of QSLT for open quantum systems as
\begin{equation}\label{a6}
\tau \geq \frac{\left|f\left(\tau+\tau_{D}\right)-1\right| tr \left(\rho_{\tau}^{2}\right)}{\overline{\sum_{i=1}^{n} \sigma_{i} \rho_{i}}},
\end{equation}
where where $\sigma_i$ and $\rho_{i}$ are the singular values of $\mathcal{L}_{t}(\rho_{t})$ and $\rho_{\tau}$, respectively and $\overline{\square}=\frac{1}{\tau_{D}} \int_{\tau}^{\tau + \tau_{D}} \square dt$. In a similar way, the MT bound of QSL-time for open quantum systems is given by
\begin{equation}\label{a7}
\tau \geq \frac{\vert f( \tau + \tau_D ) -1 \vert tr (\rho_{\tau}^{2})}{\overline{ \sqrt{\sum_{i=1}^{n} \sigma_{i}^{2}}}}.
\end{equation}
the comprehensive Bound of QSLT for open quantum system can be obtain by combining the results of MT bound and ML bound as
\begin{equation}\label{a8}
\tau_{QSL}=\max \lbrace \frac{1}{\overline{ \sum_{i=1}^{n} \sigma_{i} \rho_{i}}}, \frac{1}{\overline{ \sqrt{\sum_{i=1}^{n} \sigma_{i}^{2}}}} \rbrace \times \vert f( \tau + \tau_D ) -1 \vert tr (\rho_{\tau}^{2}).
\end{equation} 
It has been shown that  the ML bound of the QSLT is tighter than MT bound for open quantum systems\cite{Zhang}. It is worth noting that the QSLT is always smaller than driving time $\tau_D$. QSLT have an inverse relation with the speed of the quantum evolution. It means that when the QSLT reduces the evolution of the  dynamical process  exhibits a speed-up and when QSLT increases the evolution of the  dynamical process  shows a speed-down.
\section{Model}\label{sec2}
The system consists of a two-level atom coupled to a damping channel.  The time evolution of such an open quantum system under Homodyne-based feedback control is defined via the following time-dependent master equation as \cite{Rao}.
\begin{equation}\label{a9}
\frac{d \rho}{d t}=-i\left[\frac{1}{2} \omega \sigma_{z}+\frac{1}{2}\left(\sigma_{+} F+F \sigma_{-}\right), \quad \rho\right]+D\left[\sqrt{\gamma} \sigma_{-}-i F\right] \rho
\end{equation}
where 
\begin{equation}\label{a10}
F=\lambda\left(\sigma_{x} \sin \alpha+\sigma_{y} \cos \alpha\right),(\alpha \in[-\pi, \pi]),
\end{equation}
is the feedback Hamiltonian, $\sigma_{k}(k=x, y, z)$ are Pauli operators, $\omega$ is the transition frequency between the two levels, $\lambda$ is the feedback coefficient and $\gamma$ is dissipation coefficient. From feedback Hamiltonian, it is obvious that the control can be done along $x$ or $y$
directions. By starting the feedback, the evolved density matrix will be obtained as follows 
\begin{equation}\label{a11}
\rho(\varphi)=\left(\begin{array}{ll}
{\rho_{11}(t)} & {\rho_{12}(t)} \\
{\rho_{21}(t)} & {\rho_{22}(t)}
\end{array}\right)
\end{equation}
with
\begin{eqnarray}\label{a12}
\rho_{11}(t)&=&\mu(t) \rho_{11}(0)+\nu(t) \rho_{22}(0), \nonumber \\
\rho_{22}(t)&=&(1-\mu(t)) \rho_{11}(0)+(1-\nu(t)) \rho_{22}(0), \nonumber \\
\rho_{12}(t)&=&\xi(t) \rho_{12}(0)+\eta(t) \rho_{21}(0), \nonumber \\
\rho_{21}(t)&=&\eta^{*}(t) \rho_{12}(0)+\xi^{*}(t) \rho_{21}(0),
\end{eqnarray}
where 
\begin{eqnarray}\label{a13}
\mu(t)&=&\frac{1}{\Gamma}\left[\lambda^{2}+\left(\lambda^{2}+2 \lambda \sqrt{\gamma} \cos \alpha+\gamma\right) e^{-\Gamma t}\right], \nonumber \\
\nu(t)&=&\frac{1}{\Gamma} \lambda^{2}\left(1-e^{-\Gamma t}\right), \nonumber \\
\xi(t)&=&\frac{1}{\Delta} e^{-\frac{1}{2} \Gamma t}\left[\Delta \cosh \left(\frac{\Delta}{2} t\right)-i 2 \lambda \sin \alpha \sinh \left(\frac{\Delta}{2} t\right)\right], \nonumber \\
\eta(t)&=&-\frac{2}{\Delta} \lambda e^{i \alpha}\left(\sqrt{\gamma}+\lambda e^{i \alpha}\right) e^{-\frac{1}{2} \Gamma t} \sinh \left(\frac{\Delta}{2} t\right),\nonumber \\
\Gamma&=&2 \lambda^{2}+2 \lambda \sqrt{\gamma} \cos \alpha+\gamma, \nonumber \\
\Delta&=&2 \sqrt{\lambda^{2}\left(\lambda^{2}+2 \lambda \sqrt{\gamma} \cos \alpha+\gamma\right)-(\omega+\lambda \sin \alpha)^{2}}.
\end{eqnarray}
Here we choose the initial state $\vert \psi \rangle$ in the following form
\begin{equation}
|\psi\rangle=\cos \theta|0\rangle+ e^{i \chi} \sin \theta|1\rangle,
\end{equation}
where $\theta \in[0, \pi / 2]$ and $\chi \in[0, \pi]$.
\section{Result and discussion}\label{sec3}
We can now determine the QSLT for an open quantum system under Homodyne-based feedback control. Under Homodyne feedback control and by placing the value of $\alpha$ in the above equations the evolved density matrix can be obtained exactly. Depending on how the alpha coefficient is chosen, we encountered three positions as follows:
\subsection{The feedback control with $\alpha=0$}
In this situation the feedback control is applied in $y$ direction. Here we choose $\alpha=0$. So, From Eq.\ref{a10} we have 
\begin{equation}\label{a14}
F_x=0, \quad F_y= \lambda \sigma_y.
\end{equation}
In Fig.(\ref{Fig1}), the QSLT is plotted as a function of initial time $\tau$ for different value of  feedback coefficient when $\alpha=0$. As can be seen, the QSLT is decreased by increasing feedback coefficient. By increasing the feedback coefficient the QSLT reaches to zero at earlier time. In other words due to Homodyne-based feedback  control in $y$ direction, the speed of the quantum evolution increases. 
\subsection{The feedback control with $\alpha=\pi/4$}
In this situation the feedback control is applied in $xy$ direction. Here we choose $\alpha=\pi/4$. So, From Eq.\ref{a10} we have 
\begin{equation}\label{a15}
F=\frac{\lambda}{\sqrt{2}}\left(\sigma_{x} +\sigma_{y} \right).
\end{equation}
Fig.(\ref{Fig2}) shows the effect of the Homodyne-based feedback control on QSLT when $\alpha=\pi/4$. As can be seen from Fig.(\ref{Fig2}), the QSLT is decreased by increasing feedback coefficient. In other words, as the feedback coefficient increases, the QSLT will be zero at earlier initial time. We can say that the Homodyne-based feedback control will speed up the quantum evolution.
\subsection{The feedback control with $\alpha=\pi/2$}
In this section, we will study the behaviors of QSLT under Homodyne-based feedback control in $x$ direction with $\alpha=\pi/2$. So, we have
\begin{equation}\label{a16}
F_x=\lambda \sigma_x, \quad F_y= 0.
\end{equation}
In Fig.(\ref{Fig3}), we investigate the influence of Homodyne-based feedback control in $x$ direction on QSLT. Similar to the two previous cases, it can be seen here that the QSLT is decreased by increasing feedback coefficient. So here we also see that the speed of evolution of the open quantum system increases as the feedback coefficient increases. 
\section{Conclusion}\label{sec4}
In this work, we have studied the influence of three different constant feedback operator types on QSLT. We have shown how the Homodyne-based feedback control can affect the QSLT of open quantum systems. We showed that for all three cases the QSLT is decreased by increasing feedback coefficient. As the feedback coefficient increases, the QSLT will be zero at earlier initial time. In other words, we can say that the Homodyne-based feedback control will speed up the quantum evolution. So, we can use Homodyne-based feedback control to control the speed of evolution of open quantum systems.


\end{document}